\documentstyle[twocolumn,aps]{revtex}
\begin{document}
\newcommand{\beq}{\begin{equation}}
\newcommand{\eeq}{\end{equation}}

\title{Stability Threshold as a Selection Principle for Protein Design}
\author{Michele Vendruscolo$^1$, 
        Amos Maritan$^{1}$, and Jayanth  R. Banavar$^2$}
\address{$^1$ Istituto Nazionale per la Fisica della Materia (I.N.F.M.) and
International School for Advanced Studies (S.I.S.S.A.),
Via Beirut 2-4, 34014 Trieste, Italy}
\address{$^2$ Department of Physics and Center for Materials Physics,
104 Davey Laboratory, The Pennsylvania State University,
University Park, PA16802}

\address{
\centering{
\medskip\em
{}~\\
\begin{minipage}{14cm}
{}~~~ 
The sensitivity of the native
states of protein-like heteropolymers 
to mutations modelled as perturbations in the interaction potential
between amino acids is studied.  The stability threshold against mutations 
is shown to be zero for random heteropolymers  on a lattice in two dimensions,
whereas a design procedure 
modelling evolution produces a non-zero threshold.
We introduce an evolution-like protein design procedure 
based on an optimization of the stability threshold that is shown to naturally 
ensure thermodynamic stability as well.
{}~\\
{}~
\end{minipage}
}}
\maketitle

\vspace{1cm}

Natural proteins are made up of sequences of amino acids that fold
rapidly into specific compact structures corresponding to minimum free
energy states called native 
states~\cite{a73,sg89,cd93,sg93,pgt94,ssk94,pgt95,wot95,bosw95,kt96}.
The structure of the native state
conformation controls the functionality of the protein.  
Because the number of possible random amino acid sequences
and the number of possible conformations are gigantic,
an important issue is an understanding of the selection principles
that apply to protein sequences and/or native state structures.

The two key ingredients of evolution are diversity (afforded by the
availability of 20 amino acids) and stability (a functionally useful
sequence should not be mutated away). 
The stability of the occupancy of the native state on increasing the temperature
(i.e. the thermodynamic stability) has been argued to be a characteristic 
of a good folder~\cite{sg93,pgt94,dk96,svmb96}.  
Here, we consider a different kind of stability against
mutations of the sequence or equivalently perturbations in the effective
interaction potential between amino acids.
We demonstrate that the two types of stabilities are related in the sense that
each one implies the other.
We model the mechanism of evolution
through natural selection in proteins and discuss its
implication in the protein design problem.
We show that the native states of
random heteropolymers are not stable against mutations, whereas
sequences {\em designed to be thermodynamically stable}
are characterized by a non-zero stability threshold.
Conversely, an evolution-like design scheme that attempts to maximize 
the stability threshold is shown to lead to greater thermodynamic stability
as well.
Our work provides a characterization
of the ``twilight zone'' and  the observation that 
proteins form families according to the spatial conformation of their native
states~\cite{ojt94}. 
As suggested  by
Li {\em et al.}~\cite{lhtw96}, structure selection is
an appealing complementary view to sequence selection.   We show, however,
that selection processes involving sequences could be as important
as structure selection.
The effects
of destabilizing factors such as variations in the denaturant concentration
and site directed mutagenesis on the kinetics of folding have
been recently investigated in Refs. \cite{o96}.
A careful analysis of such perturbations has proven helpful
in the clarification of the folding funnel in terms of collective coordinates
\cite{wot95,bosw95,o96}.

We consider self-avoiding chains of $N$ monomers
on a 2D square lattice.
The Hamiltonian is 
\begin{equation}
H_s(\Gamma ) = \sum _{ij} B_{i,j} \delta ({\bf r}_i - {\bf r}_j ),
\end{equation}
where $B_{i,j}$ is the coupling between monomer $i$ and $j$
and $\delta$ is nonzero (and equal to 1) only if ${\bf r}_i$
and ${\bf r}_j$ are are adjacent sites on the lattice and $i$ and
$j$ are not next to each other in sequence.  
This hamiltonian is well known in protein modelling~\cite{sg89,ssk94}. 
For a given sequence (and 
$B_{i,j}$) with $N \le 25$, we enumerate the energies of all 
possible conformations
and are able to determine the native state (ground state) conformation
exactly.

We consider two  versions
of the model.  In the first (which we will call the $B_{i,j}$ model), 
the $N$ monomers are assumed to be distinct.
The $B_{i,j}$ matrix is symmetric and has  $N(N+1)/2$ elements.  
In order to obtain a random heteropolymer, these elements
are drawn from a Gaussian distribution with mean value -2 and 
variance 1.
Effectively, the matrix $B$ represents a certain sequence. 
The model is identical to that studied by
Dinner {\em et al.}~\cite{ssk94,dsks94,cvb96}. The random contact energies
are in approximate correspondence to a more realistic
parametrization of the contact energies
by Miyazawa and Jernigan~\cite{mj85,mj96} or by Kolinski, Godzik
and Skolnik~\cite{kgs93}.
In order to model evolved sequences with a large stability gap~\cite{bosw95}, 
we
follow the rank-ordered 
procedure outlined by Shrivastava {\em et al.}~\cite{svcmb95} of shuffling the
$B_{i,j}$ entries to  assign the most favorable attractive interactions 
to the native contacts of the maximally compact 
native fold chosen as fixed target.

In the second model (denoted as the MJ model), each monomer is chosen
to represent one of the twenty amino acids with the interactions determined
by Miyazawa and Jernigan~\cite{mj85,mj96}.  A random sequence
would correspond to a random choice of the amino acids.  In order to mimic
the relevant feature of sequences selected by evolution
it is no longer possible to follow the rank ordering 
procedure because the $B_{i,j}$ entries cannot be shuffled at will.
Instead, after having fixed a target fold \cite{average},
we have used a recently proposed protein design procedure~\cite{svmb96}
entailing an optimization scheme in sequence space
which allows one to obtain sequences with a desired 
native state conformation and a required measure of thermodynamic stability,
enforced by fixing a ``design temperature'' $T_d$
and selecting those sequences with $T_f > T_d$.

Our calculations begin with the selection of two 
distinct interaction matrices which we shall call $B$ and $C$.
We shall consider 4 choices for $B$:  the random and the evolved
$B_{i,j }$ and MJ models. $C$ is chosen randomly.
The ground states of the $B$ and $C$ sequences are
generally distinct.  We now consider mutations of the sequence along
a trajectory parametrized by a mixing coefficient $a \in [0,1]$ 
that changes the interaction matrix
from $B$ to $C$ \cite{other_def}:
\begin{equation}
B_a = (1-a)B + aC  .
\end{equation}
The coefficient $a$ is a measure of the 
distance in sequence space between $B$ and $B_a$.
This is a quite general perturbation of which the natural occurring ones
are a subset. 
The structural similarity of the ground state conformations of these two
sequences is given by the normalized distance 
$\Delta(a) = d(B_a,B)/d(C,B)$, where the distance $d(X,Y)$ is defined by
\begin{equation}
d(X,Y) =\sqrt{\sum_{i,j=1}^{N} (r_{i,j}-r_{i,j}')^2}
\end{equation}
where $r_{i,j}$ and $r_{i,j}'$ are the Euclidean distances
between amino acids $i$ and $j$ in the the two 
native states of sequences $X$ and $Y$ respectively.
Note that $\Delta$ has been
normalized so that it is 1 when $a=1$, as long as the ground states of
$B$ and $C$ are distinct.

Our primary probe of the stability to mutations is via a study of the dependence
of $\Delta$ on $a$.  Qualitatively similar trends are found for both
the models -- the signature of the selection in sequence
space is in the quite distinct behavior of random and evolved sequences.
A summary of our results for the behavior of the average
$\Delta$ as a function of $a$
for $N=16$ is shown in Fig. 1.  
The curves have been obtained as an average
over 1000 realizations of independently chosen $B$ and for each
of them over 1000 realizations of $C$ for 
the $B_{i,j}$ model and over 10 realizations of $B$ and for each of
them over 1000 realizations of $C$ for the MJ model.
The average stability threshold is zero for random heteropolymers~\cite{gas96} 
and is distinctly non-zero for the evolved cases.  

Furthermore, 
the stability threshold goes up with the overall
thermodynamic stability  -- in Figure 1, for the MJ model, the region
of stability against mutations increases with the design temperature $T_d$.
The threshold is somewhat reduced but is clearly non-zero when one considers
rank ordered $B_{i,j}$ sequences that have native states in conformations
that are not maximally compact.  One increasing $N$, the number of monomers,
comprising the  evolved sequences, the stable phase
along the $a$-axis increases in size along with a sharpening of the 
$\Delta-a$ curve suggestive of a sharp phase transition at the onset
of instability in the thermodynamic limit.

One may also define an individual stability threshold
$a_t(B,C)$ in the strength $a$
of the perturbation above which  $\Delta$ becomes non-zero for the first time
-- the native structure of sequence $B$ is destabilized.
Normalized probability distributions $P(a_t)$ 
of the individual stability thresholds for the random and evolved
$B_{i,j}$ models are shown in Fig. 2.  They underscore 
the  different behaviors  in the two cases.
Our results, in the random case,
are marginally related to a study of Bryngelson~\cite{b94} 
and in the evolved case, to a recent study of Pande {\em et al.} \cite{pgt95},
where the authors addressed the issue of stability of the ground state
against inaccuracies in the potential.
Bryngelson  \cite{b94} used  a mean field theory
to estimate the probability of predicting the correct structure 
of a sequence of monomers if the interatomic potential is known only to an
accuracy of $\eta$.  A non-zero $\eta$ could arise from variations in the 
solvent properties or due to the imperfect 
parametrization or determination
of the potential between amino acids or, as in our case, from mutations in
the sequences.
Pande {\em et al.} \cite{pgt95} 
showed analytically that the ground state of designed sequences 
is significantly robust against the introduction of random noise 
in the interaction matrix.

Imperfectly folded proteins within the cell are demolished
by proteolytic enzymes~\cite{creighton}.
Unfolded proteins can be present either as
a product of a destabilizing mutation
or due to a variation of the solvent properties.
Thus the folded structure must be robust against perturbations
in order to survive in the cell.
The role of sequence selection is to produce robust sequences
which  rapidly fold into 
a target conformation with a specific function.
Can one develop a design scheme  starting from existing
functional sequences and produce artificial homologues with better 
functionality?

The design scheme works as follows.
1) Select an initial random sequence. 
2) Compute its MJ matrix $B$ and its stability threshold
   by extracting a set of 100 realizations of the perturbation $C$.
3) Subject the sequence to Monte Carlo optimization procedure:
   monomers are swapped and the new sequence is accepted if its 
   stability threshold is increased.
4) After 1000 such Monte Carlo steps, 
   stop the optimization and compute the folding transition temperature $T_f$
   of the resulting sequence.
As shown in Fig. 3,  $T_f$ averaged over sequences
correlates well with the threshold.

We turn now to a recent study of Li {\em et al.}~\cite{lhtw96}
who suggested 
that certain  highly designable structures that are the unique native states 
of a large number of sequences are special in that they are thermodynamically
more stable than other structures and are stable against mutations in the 
sequence.  Their study was of chains comprising 27 monomers made up of two
kinds of amino acids, hydrophobic (H) and polar (P), on a simple cubic lattice.
The bare values of their interaction parameters 
were expressed (after scaling and shifting) 
in convenient units so that 
E(P,P)=0, E(H,P)=-1 and E(H,H)=-2.3.
Our own studies of the identical model (without an overall attractive
shift in the bare interaction
parameters that would promote maximally compact structures)
necessitate the consideration of non-maximally compact conformations as well.
We have studied a two dimensional version of the model with the  
bare unshifted 
interaction parameters considered by Li {\em et al.} and with 16 monomers.  
As suggested by them, we find 
that the conformations come with varying degrees of designability.
There are many conformations that only a few sequences have as a unique 
native state while there are few that are the native states of a large
number of sequences.
Fig. 4 shows a plot of the thermodynamic stability as measured by
a Z score~\cite{ble91} (see figure caption for a definition) versus 
the number of sequences that design the structure.  The three curves
correspond to the highest, the mean and the lowest Z score.
There is no evidence of a jump in the thermodynamic stability
beyond a certain value of $N_s$ as found  by Li {\em et al.} for the
case in which all native states are necessarily maximally compact.
Indeed, for
a given structure, there are variations in the stability on tuning the 
sequences (as in the MJ curves in Fig. 1).
Strikingly,  highly designable conformations in 2D
are always maximally compact. 
In 3D,  with the same choice of unshifted interaction parameters,
while globular structures with no holes, resembling real protein structures,
are generally the ground state of HP sequences, 
these structures are not necessarily maximally compact.
These findings show that, at least for
interactions that do not always lead to maximally compact native states, the
selection process primarily involves the sequences and not the structure.
The present study has concentrated on the thermodynamic effects of mutations.
The two-way link between resistance against mutations and thermodynamic
stability demonstrated here
could also have ramifications on the  dynamics of folding,
for example, in the key role that
specific amino acids have in nucleation mechanisms of folding
\cite{o96,sap96}.

We are grateful to Hao Li and Chao Tang for useful discussions.
This work was supported by 
NASA, NATO, NSF, The Petroleum Research Fund administered
by the American Chemical Society and the Center for Academic Computing at
Penn State.

%%%%%%%%%%%%%%%%%%%%%%  BIBLIOGRAPHY  %%%%%%%%%%%%%%%%%%%%%%%%%%%%%%%%%%%

%%%%%%%%%%%%%%%%%%%%%%  FIGURES  %%%%%%%%%%%%%%%%%%%%%%%%%%%%%%%%%%%%%%%%

%%%%%%%%%%%%%%%%%%%%%%  FIG 1

\begin{figure}
\caption{
Average structural similarity $\Delta$ as a function of the 
perturbation $a$. 
The non evolved and evolved cases for both the models used are shown.
For the $B_{i,j}$ model the averages have been taken over 1000 realizations
of $B$ and for each of them over 1000 realizations of $C$.
For the MJ model we averaged over 10 realizations of $B$ and for each
of them over 1000 realizations of $C$. 
In our design procedure  for the MJ model,
only $B$ sequences with a 
folding transition temperature $T_f$ greater than the indicated value 
of the design temperature $T_d$ (0.0,1.0, and 1.2, respectively) 
were
considered. $T_f>0$
indicates no selection.
}
\end{figure}

%%%%%%%%%%%%%%%%%%%%%%  FIG 2

\begin{figure}
\caption{
The normalized probability distributions $P(a_t)$
of the individual stability thresholds $a_t$
for the non evolved and
evolved $B_{i,j}$ case. The statistics are the same as in Fig. 1.
}
\end{figure}

%%%%%%%%%%%%%%%%%%%%%%  FIG 3

\begin{figure}
\caption{
Folding transition 
temperature $T_f$ as a function of the stability threshold $a$
in the MJ model. The sequences have been  designed through
threshold optimization.
}
\end{figure}

%%%%%%%%%%%%%%%%%%%%%%  FIG 4

\begin{figure}
\caption{
The $Z$ score plotted against $N_S$ for the 2D version of the 3D model
considered by Li {\em et al.}\protect\cite{lhtw96}. 
The $Z$ score is the difference between the average energy $\langle E \rangle$
of the compact conformations and the ground state energy, $E$, 
scaled by the dispersion $\sigma$,
$Z= (\langle E \rangle)-E/\sigma$.
$\langle E \rangle$ and $\sigma$ were calculated as averages 
over all conformations
with 7 or more contacts,
which are those competing to be the native state.  
Maximally compact conformations have nine contacts.
The $Z$ score is a measure of the thermodynamic stability of the native state.
Similar trends are observed on considering the energy gap instead of
the $Z$ score.
}
\end{figure}


\begin{thebibliography}{99}
\bibitem{average}
For both the $B_{i,j}$ and MJ models, 
we have repeated the analysis for three different target folds,
with no significant difference in the results.
\bibitem{other_def}
In the case of the $B_{i,j}$ model, a more appropriate definition
would be $B_a=\sqrt{1-a^2}B+aC$. With such a definition, the variance
of $B_a$ would be unaffected and therefore the scale of the folding
transition temperature would be unchanged.
See Refs. \cite{sg89,pgt95}.
\end{thebibliography}
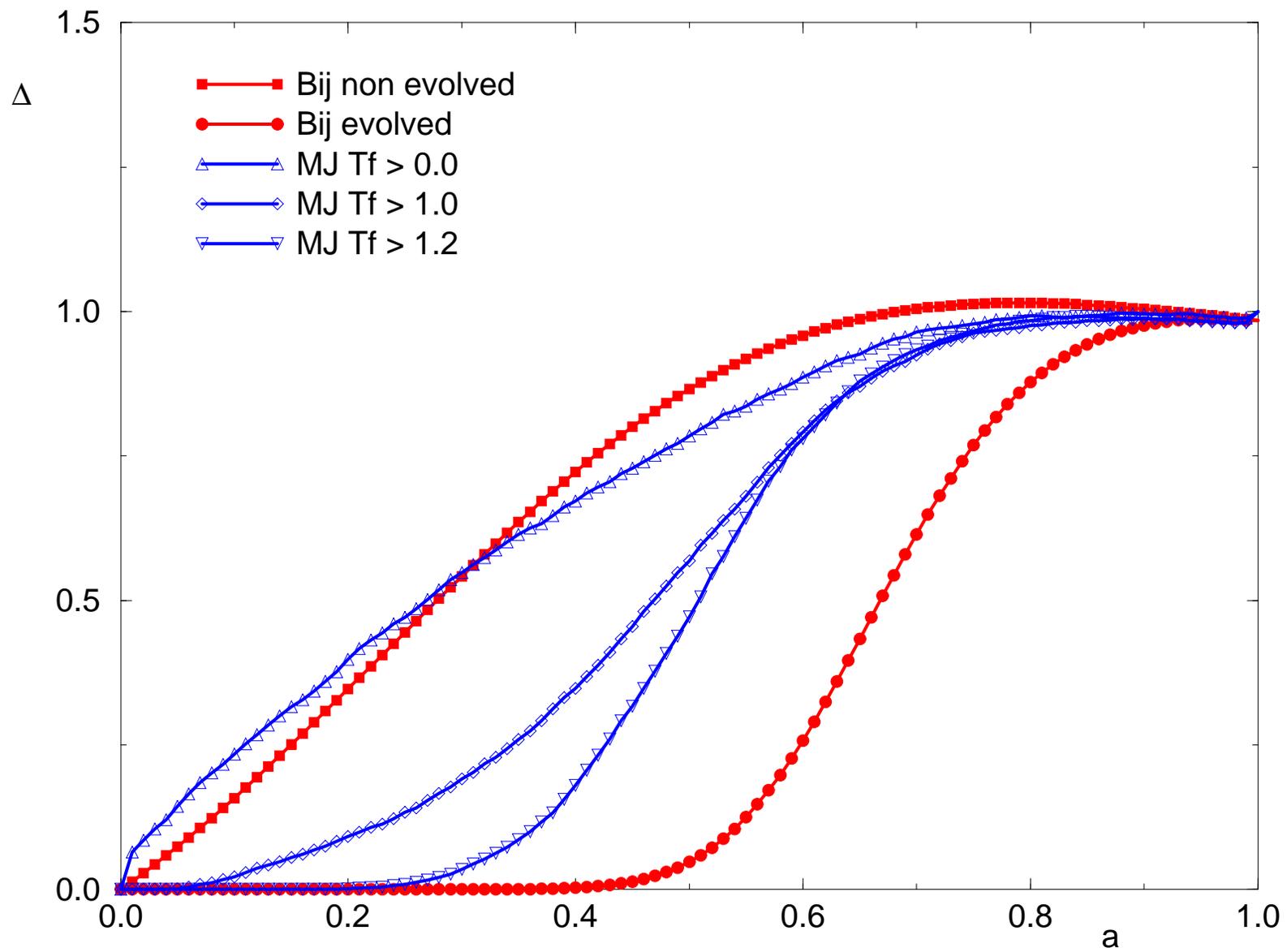
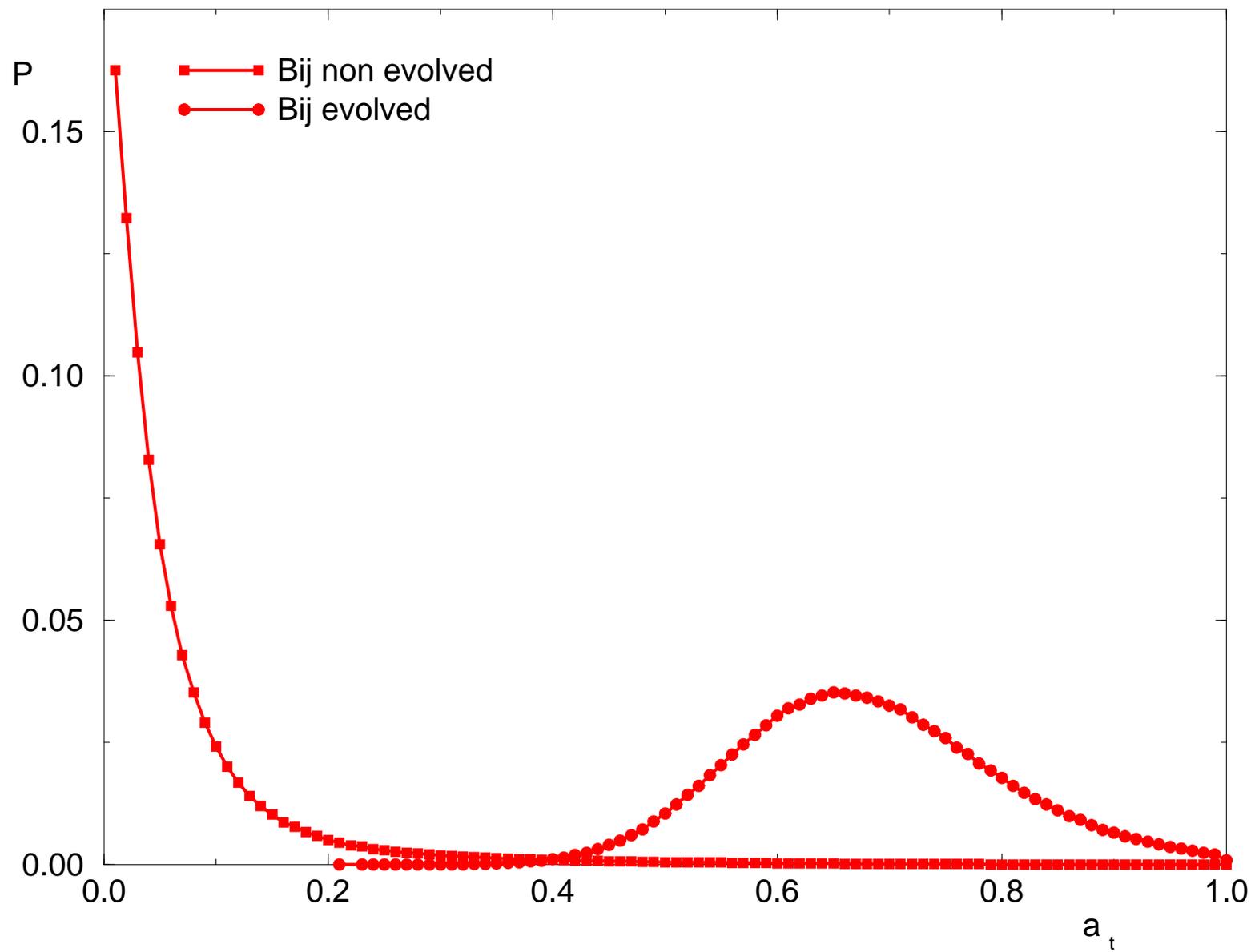
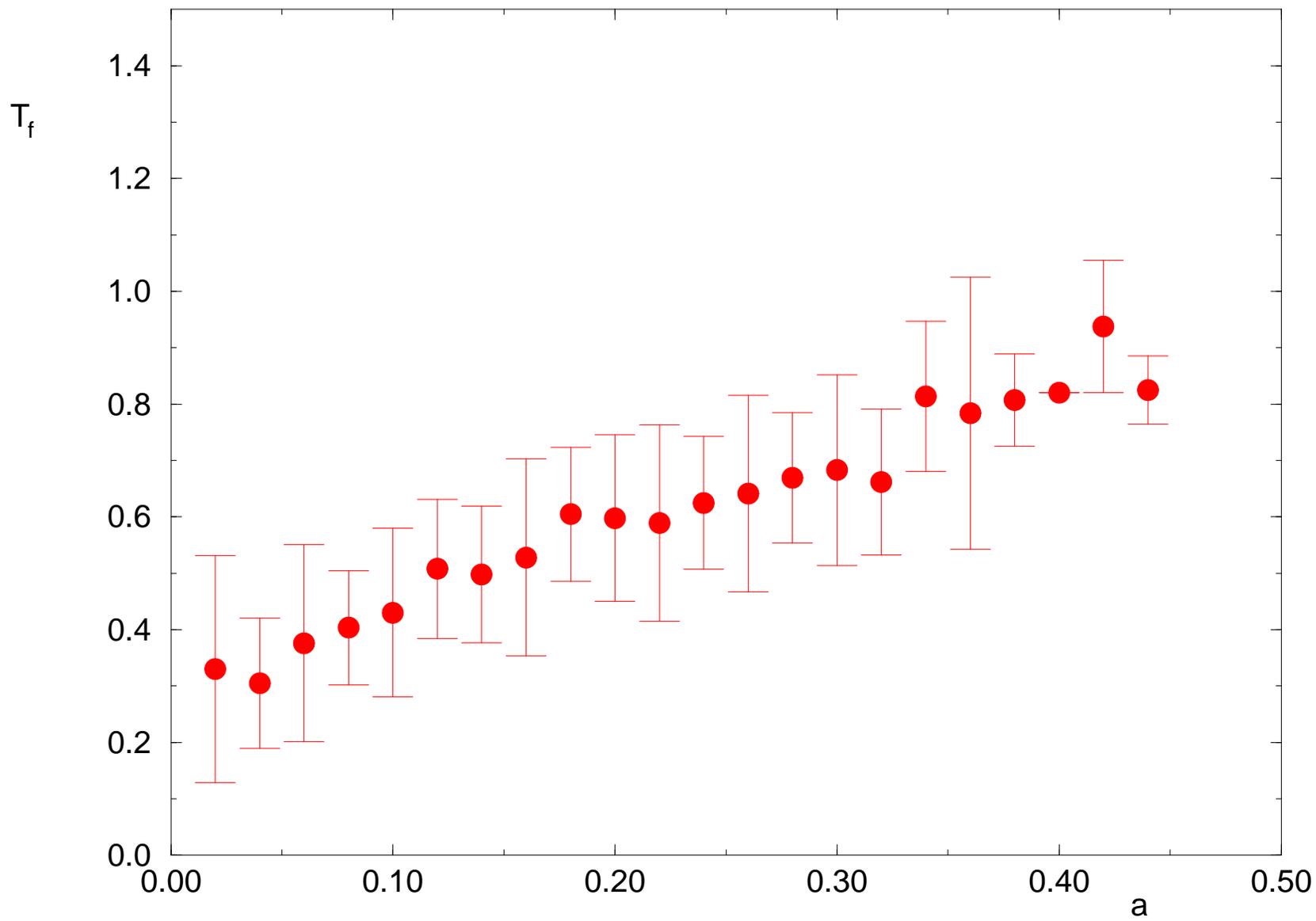
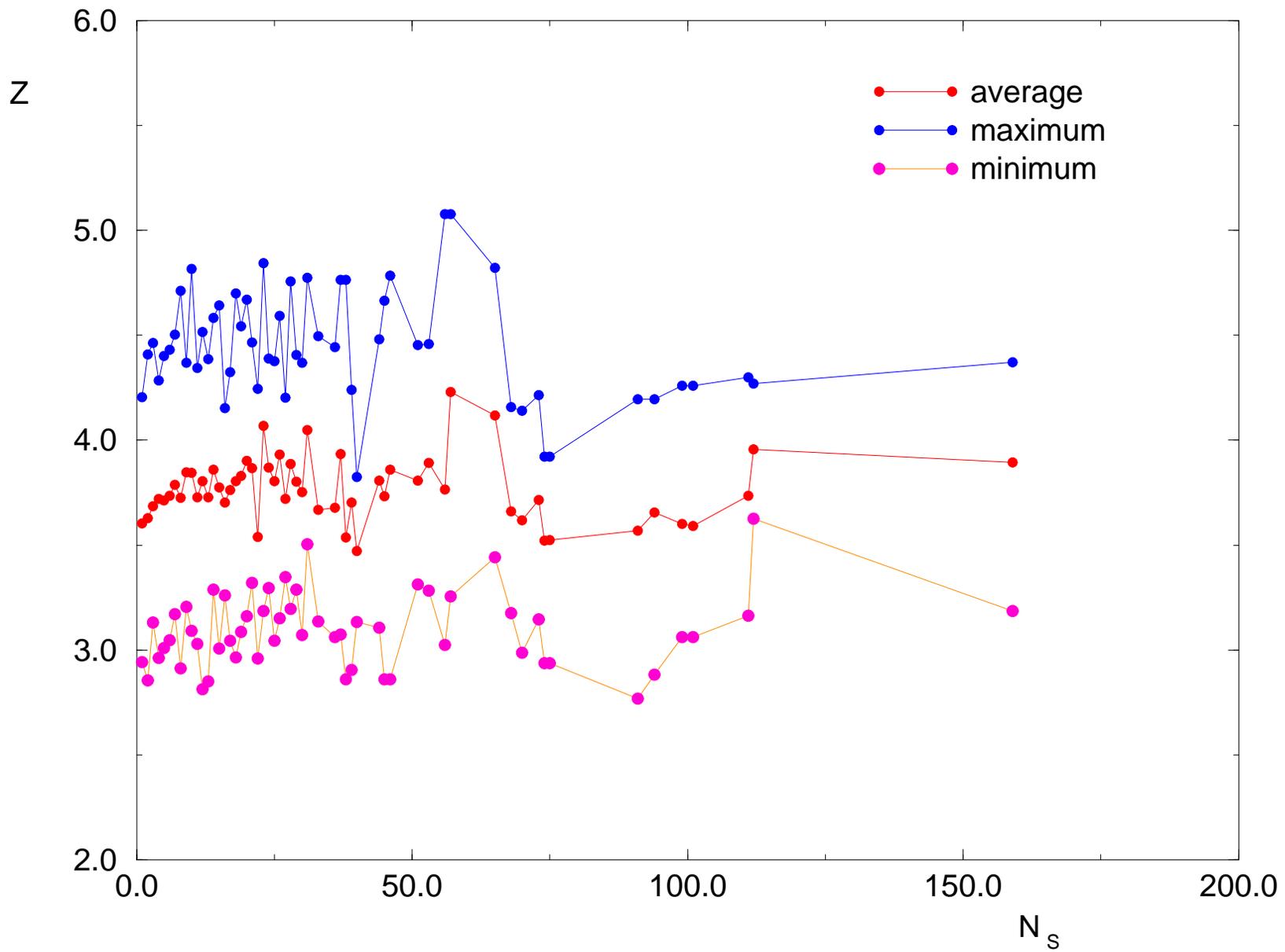
\end{document}